\documentclass[aps,preprintnumbers,superscriptaddress]{revtex4}
\usepackage{epsfig}
\usepackage{psfrag}
\usepackage{amsfonts}
\usepackage{graphicx}
\usepackage{dcolumn}
\usepackage{bm}
\usepackage{amsmath}
\usepackage{amssymb}
\usepackage{cancel}
\usepackage{stackrel}
\usepackage{esint}
\usepackage{slashed}
\usepackage{mathtools}
\begin{document}

\title{The HTL resumed  propagators in the light cone gauge}

\author{Qi Chen}
\email{qichen@mails.ccnu.edu.cn} \affiliation{Key Laboratory of Quark
and Lepton Physics (MOE), Central China Normal University, Wuhan
430079, China.}

\author{De-fu Hou}
\email{hdf@mail.ccnu.edu.cn} \affiliation{Key Laboratory of Quark
and Lepton Physics (MOE), Central China Normal University, Wuhan
430079, China.}

\begin{abstract}

\indent
The expression of the HTL resumed  gluon propagator in the light cone gauge is derived.
In the real time mechanism,
using the Mandelstam-Leibbrant prescription of $(n\cdot K)^{-1}$,
we calculate the transverse and longitudinal parts of the gluon HTL self-energy  and prove the transverse and longitudinal parts do not have  divergence.
We also calculate the quark self energy in the HTL approximation,
and find it gauge independent. We analytically calculate the damping rates of the hard  quark and gluon with this HTL resumed gluon propagator.
\end{abstract}

\maketitle
\section{Introduction}

\indent The bare QCD perturbative theory breaks down at high temperature.
There are some serious problems of gauge theories at finite temperature,
such as IR singularity and gauge dependent results,
when the bare propagators (vertices) are used.
The HTL (Hard Thermal Loop) resumed propagators have been developed by Braaten and Pisarski \cite{BP1990}.
 Some gauge independent physical quantities are given with the HTL resumed propagators in the calculation,
other than  the bare propagators at finite temperature.
If the momentum of the propagators are soft at finite temperature,
we should use the HTL  resumed  propagators.
High order loop HTL diagrams can give a low order contribution in the coupling constant at finite temperature,
which should be resumed. Because of the HTL resummation,  the medium effects are taken into account,
such  as the Debye screening caused by the color charges of the QGP.
The HTL resummation technique represents a great progress compared to the bare perturbation theory at finite temperature.

The light cone gauge is  one of non-covariant  and physical gauges \cite{L},\cite{BNS1991}, and is also ghost-free.
When  the multiple gluon emission is calculated in the light cone gauge, because  the  interference terms among different tree diagrams do not contribute to the leading order in the process of calculating the diagram amplitude  in the light cone gauge, the differential cross section with $n$-gluon emission in the leading pole approximation has a simple ladder structure  at zero temperature \cite{CSS89}. These nice properties simplify the calculation.
 However, the light cone gauge has its disadvantage,
 such as the spurious singularity of $(n\cdot K)^{-1}$ ,
 the renormalization and so on.

In the experiment of the  Heavy Ion Collisions,
the longitudinal momentum of the generated parton  is very large, however,
the transverse momentum is very small, and it is more suitable to calculate some physical quantities in the light cone gauge with light cone variables  $K^{\mu}=(k^{+},k^{-},k_{\perp})$, than to do in the Coulomb gauge in the Minkowski space $K^{\mu}=(k^{0},\overrightarrow{k})$.  Many good theoretical works have been done in light cone gauge in vacuum  which  are well consistent with the experiment data.

Recently we can only calculate the evolution equation of  parton distribution  functions (PDFs) \cite{MSTW2009},\cite{CSS1989} and parton fragmentation  functions (FFs)\cite{MV2016},\cite{ES2017},\cite{WG2001}  by pertubative QCD, which are both non-perturbative physical quantities.  The evolution equations govern the running  of PDFs and FFs with the scale $Q$.  Correspondingly, we can  extract  PDFs and FFS from the experimental data, which are taken some fixed value for the relevant hard scale $Q$.

For some physical quantities such as the FFs \cite{OWW2003} and so on, we can extend the case of  high energy in vacuum to  the case at  high temperature.
For hard processes, we can use the bare propagators in the light cone gauge at  finite temperature.
With the  HTL resumed  gluon  propagator in  the light cone gauge,  we can consider multiple soft gluon  scattering among hard partons and hot medium in the Heavy Ion Collisions, which contains soft process on the basis of the hard process. The soft process can give  a significant correction   compared with  the hard process. This is the motivation of this paper. We  work out the transverse and longitudinal parts of the gluon HTL  self energy in the light cone gauge, derive the HTL resumed gluon  propagator and demonstrate it is gauge independent for further research in the future.

The remainder of the paper is organized as follows.
In Sec.II we review the HTL  resumed gluon propagator in the Coulomb gauge. In Sec.III the HTL resumed gluon propagator in the light cone gauge is worked out,
and then we calculate and analyze the transverse and longitudinal parts of the  gluon HTL self energy in the light cone gauge.
Via the HTL resumed gluon propagator we show the transverse and longitudinal spectral functions and the equations of the  dispersion relation.
In Sec.IV we calculate the  quark self energy in the  HTL approximation, and show the HTL resumed  quark propagator gauge independent.
In Sec.V we analytically calculate the damping rates of the hard quark and gluon with this HTL resumed gluon propagator in the light cone gauge in a particular limit, and demonstrate in general case we can have the same result in the light cone gauge and the Coulomb gauge.
Our conclusion is given in Sec.VI. We define the notation $P^{\mu}=(p^{0},\overrightarrow{p})$, etc.

\section{HTL resumed gluon propagator in the Coulomb gauge}
\indent At zero temperature, covariant gauge has a definite advantage over non-covariant gauges such as the Coulomb gauge or axial gauges.
Calculations are simplified considerably due to Lorentz invariance,
and the renormalization program can be implemented in practice only in covariant gauge.
At finite temperature, Lorentz  invariance is broken because the heat bath defines a privileged frame,
and renormalization program is of secondary importance,
so that non-covariant gauges may present useful alternatives to covariant gauge\cite{B1996}.

The HTL resumed gluon propagator has been derived  in  the Coulomb gauge \cite{We1982},\cite{ASS2010}.
The HTL gluon self energy $\Pi^{\mu\nu}(P)$ is expressed as the transverse part and the longitudinal part.
The gluon self energy  is given by
\begin{equation}\label{selfenergyCoulomb}
 \Pi^{\mu\nu}(P)=-\Pi_{T}(P)T_{P}^{\mu\nu}-\frac{1}{n_{P}^{2}}\Pi_{L}(P)L_{P}^{\mu\nu}\;,
\end{equation}
where the transverse projection tensor $T_{P}^{\mu\nu}$, the longitudinal  projection tensor $L_{P}^{\mu\nu}$, and the four vector $n_{P}^{\mu}$ are defined as
\begin{eqnarray}\label{TLCoulomb}
T_{P}^{\mu\nu}&=&g^{\mu\nu}-\frac{P^{\mu}P^{\nu}}{P^{2}}-\frac{n_{P}^{\mu}n_{P}^{\nu}}{n_{P}^{2}}\notag\;,\\
L_{P}^{\mu\nu}&=&\frac{n_{P}^{\mu}n_{P}^{\nu}}{n_{P}^{2}}\;,\notag\\
n_{P}^{\mu}&=&n^{\mu}-\frac{n\cdot P}{P^{2}}P^{\mu}\;.
\end{eqnarray}

The axial vector  is
\begin{equation}\label{axial-vector}
n_{c}^{\mu}=(n^{0},n^{1},n^{2},n^{3})=(1,0,0,0)\;,
\end{equation}
which specifies the thermal rest  frame.

The inverse propagator for general $\xi$ in the Coulomb gauge  is
\begin{equation}
\Delta_{\xi}^{-1}(P)^{\mu\nu}=\Delta^{-1}(P)^{\mu\nu}-\frac{1}{\xi}(P^{\mu}-P\cdot nn^{\mu})(P^{\nu}-P\cdot nn^{\nu})\;,
\end{equation}
where $\xi$ is a arbitrary gauge parameter.

The inverse propagator reduces in the limit $\xi \rightarrow \infty$ to
\begin{equation}
\Delta_{\infty}^{-1}(P)^{\mu\nu}=-P^{2}g^{\mu\nu}+P^{\mu}P^{\nu}-\Pi^{\mu\nu}(P)\;.
\end{equation}

$\Delta_{\infty}^{-1}(P)^{\mu\nu}$ can also be written as
\begin{equation}
\Delta_{\infty}^{-1}(P)^{\mu\nu}=-\frac{1}{\Delta_{T}(P)}T_{P}^{\mu\nu}+\frac{1}{n_{P}^{2}\Delta_{L}(P)}L_{P}^{\mu\nu}\;,
\end{equation}
where $\Delta_{T}(P)$ and$\Delta_{L}(P)$ are the transverse and longitudinal propagators:
\begin{eqnarray}
\Delta_{T}(P)&=&\frac{1}{P^{2}-\Pi_{T}(P)}\;,\notag\\
\Delta_{L}(P)&=&\frac{1}{-n_{P}^{2}P^{2}+\Pi_{L}(P)}\;.
\end{eqnarray}

The HTL resumed gluon propagator in the Coulomb gauge \cite{ASS2010} is
\begin{equation}
\Delta^{\mu\nu}_{\xi}(P)=-\Delta_{T}(P)T_{P}^{\mu\nu}+\Delta_{L}(P)n^{\mu}n^{\nu}-\xi\frac{P^{\mu}P^{\nu}}{(n_{P}^{2}P^{2})^{2}}\;.
\end{equation}

By calculating, we can find
\begin{equation}
 T_{P}^{00}=0,T_{P}^{0i}=T_{P}^{i0}=0\;.
\end{equation}
So the  HTL resumed gluon propagator in the Coulomb gauge can be simplified into
\begin{eqnarray}\label{propagator-in-Coulomb-gauge}
G_{Ret}^{00}(P)&=&\frac{1}{p^{2}+\Pi_{L}(P)}\notag\;,\\
G_{Ret}^{ij}(P)&=&\frac{\delta^{ij}-\hat{p}^{i}\hat{p}^{j}}{P^{2}-\Pi_{T}(P)}\;,
\end{eqnarray}
where $\hat{\overrightarrow{p}}$ is a unit vector on the direction of $\overrightarrow{p}$, $\hat{\overrightarrow{p}}=\frac{\overrightarrow{p}}{|\overrightarrow{p}|}
 $ and $\hat{\overrightarrow{p}}=(\hat{p}^{1},\hat{p}^{2},\hat{p}^{3})$.

The  longitudinal and  transverse gluon HTL self energy \cite{T2000} are
\begin{eqnarray}\label{TL-parts}
\Pi_{L}(P)&=&m_{D}^{2}\left[1-\frac{p_{0}}{2p}\ln|\frac{p_{0}+p}{p_{0}-p}|+i\pi\frac{p_{0}}{2p}\theta(p^{2}-p_{0}^{2})\right]\notag\;,\\
\Pi_{T}(P)&=&\frac{m_{D}^{2}}{2}\frac{p_{0}^{2}}{p^{2}}\left[1-(1-\frac{p^{2}}{p_{0}^{2}})\frac{p_{0}}{2p}\left[\ln|\frac{p_{0}+p}{p_{0}-p}|-i\pi\theta(p^{2}-p_{0}^{2})\right]\right]\;,
\end{eqnarray}
where the gluon screening mass $m_{D}^{2}=\frac{1}{3}(C_{A}+\frac{1}{2}N_{f})g^{2}T^{2}$ and $\theta(p^{2}-p_{0}^{2})$ is the step function.

The imaginary parts in the above equations correspond to the Landau damping,
which means that one particle is emitted from the thermal medium and absorbed by the medium.

In the static limit, $p_{0}\rightarrow 0$, the longitudinal HTL self energy
\begin{equation}
\Pi ^{L}_{R}(p_{0}\rightarrow 0,p)=m_{D}^{2}\;,
\end{equation}
which means the Debye screening of the gluon in the plasma.

However, in the static limit, the transverse HTL self energy
\begin{equation}
\Pi ^{T}_{R}(p_{0}\rightarrow 0,p)=0\;,
\end{equation}
which shows no static magnetic screening.

The self-energy tensor $\Pi^{\mu\nu}$ is  symmetric in $\mu$ and $\nu$ and satisfies
\begin{eqnarray}\label{relationTL}
&&P_{\mu}\Pi^{\mu\nu}(P)=0\notag\;,\\
&&g_{\mu\nu}\Pi^{\mu\nu}(P)=-2\Pi_{T}(P)-\frac{1}{n_{P}^{2}}\Pi_{L}(P)=-m_{D}^{2}\;.
\end{eqnarray}

\section{The HTL resumed gluon propagator in the light cone gauge}
\label{sec3}

\indent In this section, we derive the HTL resumed gluon  propagator in the light cone gauge,
and compute the transverse and longitudinal parts of gluon HTL  self energy in the real time formalism.
In the static limit, we discuss $\Pi^{00}(P)$ in the light cone gauge and the Coulomb gauge.
We  obtain the pole terms and the cut terms of the transverse and longitudinal spectral functions.

The light cone gauge is one of axial type gauges and non-covariant gauges \cite{L},\cite{BNS1991},
\begin{equation}
n_{l}^2=0,n_{l}\!\cdot\! A=0\;.
\end{equation}

The axial vector in the light cone gauge is
\begin{equation}\label{lcg-axial-vector}
n_{l}^{\mu}=(n^{0},n^{1},n^{2},n^{3})=(\frac{\sqrt{2}}{2},0,0,-\frac{\sqrt{2}}{2})\;.
\end{equation}

The bare gluon propagator in the light cone gauge is
\begin{equation}\label{bare-gluon-propagator}
\frac{i(-g^{\mu\nu}+\frac{n^{\mu}K^{\nu}+n^{\nu}K^{\mu}}{n\cdot K})}{K^{2}+i\epsilon}\;.
\end{equation}

Here we use the Mandelstam-Leibbrandt (ML) prescription of $(n\cdot K )^{-1}$ instead of the usual principal-value prescription,
\begin{equation}\label{MLprescription}
\frac{1}{n\!\cdot\! K}=\frac{n^{*}\!\cdot\! K}{n\!\cdot\! K n^{*}\!\cdot\! K+i\epsilon}=\frac{1}{n\!\cdot\! K+isgn(n^{*}\!\cdot\! K)\epsilon}=\frac{n_{0}k_{0}+\overrightarrow{n}\!\cdot\!\overrightarrow{k}}{(n_{0}k_{0})^{2}-(\overrightarrow{n}\!\cdot\!\overrightarrow{k})^{2}+i\epsilon}\;.
\end{equation}
where $n_{l}^{*\mu}=(n^{0},n^{1},n^{2},n^{3})=(\frac{\sqrt{2}}{2},0,0,\frac{\sqrt{2}}{2})$.

The usual principal-value prescription of $(n\cdot K)^{-1}$ leads to some serious problems,
such as violating power counting and other basic criteria,
when we calculate the integral of the loop diagram\cite{L}.

In the real time mechanism, the time  of the field goes from $t=0$ to $t=-i\beta$. The contour can be deformed in order to include the real time axis by going first from $t=0$ to $t=\infty$ above the real time axis and then back to $t=-i\beta$ below real time axis. So we have double degrees of freedom, one exists above the real time axis and the other one exists below the real time axis.
We get the propagator in the real time formalism \cite{B1996},\cite{CSHY1985}, which is a $2\times 2$ matrix,
\begin{equation}\label{RTFpropagator}
  \Delta(K)=\begin{pmatrix}\Delta_{11} & \Delta_{12}\\
\Delta_{21} & \Delta_{22}
\end{pmatrix}=\begin{pmatrix}\frac{1}{K^{2}-m^{2}+i\epsilon} & 0\\
0 & \frac{-1}{K^{2}-m^{2}-i\epsilon}
\end{pmatrix}-2\pi i\delta(K^{2}-m^{2})\begin{pmatrix}n_{B}(k_{0}) & \theta(-k_{0})+n_{B}(k_{0})\\
\theta(-k_{0})+n_{B}(k_{0}) & n_{B}(k_{0})
\end{pmatrix}\;.
\end{equation}

The Bose-Einstein distribution function $n_{B}(k_{0})$is
\begin{equation}
 n_{B}(k_{0})=\frac{1}{e^{|\frac{k_{0}}{T}|}-1}\;.
\end{equation}

For fermions, we have

\begin{small}
\begin{eqnarray}\label{RTF-q-propagator}
  &&F(K)=(\cancel{K}+m)
 \tilde{\Delta}(K)\!=\!(\cancel{K}+m)
 \begin{pmatrix}\tilde{\Delta}_{11} & \tilde{\Delta}_{12}\\
\tilde{\Delta}_{21} & \tilde{\Delta}_{22}
\end{pmatrix}\notag\\
&&=(\cancel{K}+m)
 \bigg[\begin{pmatrix}\frac{1}{K^{2}\!-\!m^{2}\!+\!i\epsilon} & 0\\
0 & \frac{-1}{K^{2}-m^{2}-i\epsilon}
\end{pmatrix}-2\pi i\delta(K^{2}-m^{2})\begin{pmatrix}-f(k_{0}) & \theta(-k_{0})\!-\!f(k_{0})\\
\theta(-k_{0})\!-\!f(k_{0}) & -f(k_{0})
\end{pmatrix}\bigg]\;,
\end{eqnarray}\end{small}

The Fermi-Dirac distribution function  $f(k_{0})$ is

\begin{equation}
  f(k_{0})=\frac{1}{e^{|\frac{k_{0}}{T}|}+1}\;.
\end{equation}

We use the Keldysh representation in real time formalism \cite{CSHY1985},\cite{K1965}.
The retarded propagator, advanced propagator and symmetric propagator for bosons are

\begin{eqnarray}
&&\Delta_{R}(K)=\Delta_{11}-\Delta_{12}=\frac{1}{K^{2}-m^{2}+isgn(k_{0})\epsilon}\notag\;,\\
&&\Delta_{A}(K)=\Delta_{11}-\Delta_{21}=\frac{1}{K^{2}-m^{2}-isgn(k_{0})\epsilon}\notag\;,\\
&&\Delta_{S}(K)=\Delta_{11}+\Delta_{22}=-2\pi i\delta(K^{2}-m^{2})[1+2n_{B}(k_{0})]\;.
\end{eqnarray}

The inverse relation for bosons is

\begin{eqnarray}\label{RTFGreen-functions}
&&\Delta_{11}=\frac{1}{2}[\Delta_{S}(K)+\Delta_{A}(K)+\Delta_{R}(K)]\notag\;,\\
&&\Delta_{12}=\frac{1}{2}[\Delta_{S}(K)+\Delta_{A}(K)-\Delta_{R}(K)]\notag\;,\\
&&\Delta_{21}=\frac{1}{2}[\Delta_{S}(K)-\Delta_{A}(K)+\Delta_{R}(K)]\notag\;,\\
&&\Delta_{22}=\frac{1}{2}[\Delta_{S}(K)-\Delta_{A}(K)-\Delta_{R}(K)]\;.
\end{eqnarray}

For fermions, in the Keldysh representation,
we only replace the Bose-Einstein distribution function $n_{B}(k_{0})$ by the Fermion-Dirac distribution function $-f(k_{0})$ in the symmetric propagator, and the retarded propagator and advanced propagator are the same as that of bosons,

\begin{equation}
\tilde{\Delta}_{S}(K)=\tilde{\Delta}_{11}+\tilde{\Delta}_{22}=-2\pi i\delta(K^{2}-m^{2})[1-2f(k_{0})]\;.
\end{equation}

The inverse relation in Eq.\eqref{RTFGreen-functions} is  also applicable for fermions.

\subsection{The gluon HTL self energy in the light cone gauge}

It has been checked by explicit computation in different gauges (covariant, Coulomb, temporal) whose axial vectors are all the same,
 $n_{c}^{\mu}=(1,0,0,0)$,
that the  gluon HTL self energy  does not depend on the choice of gauge. However,
the axial vector in light cone gauge is different, that brings about some changes.

A massive boson gives rise to the longitudinal polarization state, so that the boson self energy is separated into the longitudinal and transverse parts \cite{CSS89}.
The massive gluon self energy in the light cone gauge is made up of the transverse and longitudinal parts,
\begin{equation}\label{selfenergylgc}
\Pi^{\mu\nu}(P)=-\Big[\tilde{T}_{P}^{\mu\nu}\Pi_{T}(P)+\frac{\tilde{L}_{P}^{\mu\nu}}{n_{P}^{2}}\Pi_{L}(P)\Big]\;.
\end{equation}

The transverse  projection tensor is
\begin{equation}\label{transverse}
\tilde{T}_{P}^{\mu\nu}=g^{\mu\nu}-\frac{n^{\mu}P^{\nu}+n^{\nu}P^{\mu}}{n\cdot P}+\frac{n^{\mu}n^{\nu}P^{2}}{(n\cdot P)^{2}}\;.
\end{equation}

The longitudinal projection tensor is
\begin{equation}\label{longitudinal}
\tilde{L}_{P}^{\mu\nu}=-\left[\frac{n^{\mu}n^{\nu}P^{2}}{(n\cdot P)^{2}}-\frac{n^{\mu}P^{\nu}+n^{\nu}P^{\mu}}{n\cdot P}+\frac{P^{\mu}P^{\nu}}{P^{2}}\right]\;.
\end{equation}

The four-vector $n_{P}^{\mu}$ is
\begin{equation}
n_{P}^{\mu}=\Big(g^{\mu\nu}-\frac{P^{\mu}P^{\nu}}{P^{2}}\Big)n_{\nu}=n^{\mu}-\frac{n\cdot P}{P^{2}}P^{\mu}\;.
\end{equation}

$\tilde{L}_{p}^{\mu\nu}$ and $\tilde{T}_{p}^{\mu\nu}$ satisfy the following  relations,
\begin{eqnarray}\label{TLrelation}
\tilde{L}_{P}^{\mu\nu}\tilde{L}_{P\nu}^{\rho}&=&\tilde{L}_{P}^{\mu\rho}\;,\notag\\
\tilde{T}_{P}^{\mu\nu}\tilde{T}_{P\nu}^{\rho}&=&\tilde{T}_{P}^{\mu\rho}\;,\notag\\
\tilde{T}_{P}^{\mu\rho}\tilde{L}_{P\mu\sigma}&=&0\;.
\end{eqnarray}
These equations are also suitable for $L_{P}^{\mu\nu}$ and $T_{P}^{\mu\nu}$  in Eq.\eqref{TLCoulomb} in the Coulomb gauge.

The axial vector $n^{\mu}$ in the light cone gauge is defined in Eq.\eqref{lcg-axial-vector},
\begin{eqnarray}\label{axial-tensors}
n_{\mu}\tilde{T}_{P}^{\mu\nu}&=&0\;,\notag\\
n_{\mu}\tilde{L}_{P}^{\mu\nu}&\not=&0\;.
\end{eqnarray}
$\tilde{L}_{P}^{\mu\nu}$ and $\tilde{T}_{P}^{\mu\nu}$ are the longitudinal and transverse projection tensors with respect to the axial vector $n^{\mu}$ in the light cone gauge in Eq.\eqref{lcg-axial-vector}.
Because the axial vector in the light cone gauge in Eq.\eqref{lcg-axial-vector} is different from the  axial vector in the Coulomb gauge in Eq.\eqref{axial-vector},
so the longitudinal and transverse projection tensors $\tilde{L}_{P}^{\mu\nu}$ and $\tilde{T}_{P}^{\mu\nu}$ are different from those in the Coulomb gauge.
Finally, these differences make the expression of the HTL resumed gluon propagator in the light cone gauge changed.

The inverse propagator in the light cone gauge in the limit $\xi \to \infty$ is
\begin{eqnarray}
\Delta_{\infty}^{-1}(P)^{\mu\nu}&=&-P^{2}g^{\mu\nu}+P^{\mu}P^{\nu}-\Pi^{\mu\nu}(P)\;,\\
&=&\big[-P^{2}+\Pi_{T}(P)\big]\tilde{T}_{P}^{\mu\nu}+\big[-P^{2}+\frac{1}{n_{P}^{2}}\Pi_{L}(P)\big]\tilde{L}_{P}^{\mu\nu}\;.
\end{eqnarray}

Applying Eq.\eqref{TLrelation},\eqref{axial-tensors}, we can get the HTL resumed gluon propagator in the light cone gauge
\begin{equation}
\Delta^{\mu\nu}(P)=\frac{\tilde{T}_{P}^{\mu\nu}}{-P^{2}+\Pi_{T}(P)}+\frac{-\frac{n^{\mu}n^{\nu}P^{2}}{(n\cdot P)^{2}}}{-P^{2}+\frac{1}{n_{P}^{2}}\Pi_{L}(P)}\;,
\end{equation}
where we do not consider the terms containing the gauge parameter $\xi$.

When $\Pi_{T}(P)=0$ and $\Pi_{L}(P)=0$, the HTL resumed gluon propagator returns back to the bare gluon propagator in Eq.\eqref{bare-gluon-propagator}.

Due to the relation in Eq.\eqref{selfenergylgc},\eqref{transverse},\eqref{longitudinal},\eqref{TLrelation}, the transverse and longitudinal gluon HTL self-energies are given by
\begin{eqnarray}\label{TL}
\Pi_{T}(P)&=&-\frac{1}{2}\tilde{T}_{P\mu\nu}\Pi^{\mu\nu}(P)=-\frac{1}{2}\left[g_{\mu\nu}-\frac{n_{\mu}P_{\nu}+n_{\nu}P_{\mu}}{n\cdot P}+\frac{n_{\mu}n_{\nu}P^{2}}{(n\cdot P)^{2}}\right]\Pi^{\mu\nu}(P),\\
\frac{1}{n_{P}^{2}}\Pi_{L}(P)&=&-\tilde{L}_{P\mu\nu}\Pi^{\mu\nu}(P)=\left[\frac{P_{\mu}P_{\nu}}{P^{2}}-\frac{n_{\mu}P_{\nu}+n_{\nu}P_{\mu}}{n\cdot P}+\frac{n_{\mu}n_{\nu}P^{2}}{(n\cdot P)^{2}}\right]\Pi^{\mu\nu}(P)\;.
\end{eqnarray}
$\Pi^{\mu\nu}(P)$ is the sum of the quark loop,
the gluon loop and the gluon tadpole in the light cone gauge.
Multiply $\Pi^{\mu\nu}(P)$ by the projection tensors $-\frac{1}{2}\tilde{T}_{P\mu\nu}$ and $-\tilde{L}_{P\mu\nu}$,
and we can calculate $\Pi_{T}(P)$ and $\frac{1}{n_{P}^{2}}\Pi_{L}(P)$ in the HTL approximation.

\subsection{The quark loop of the gluon HTL self energy in the light cone gauge}

\begin{figure}
  \centering
  \includegraphics[width=4cm]{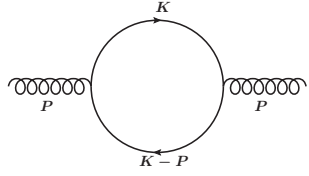}\\
  \caption{The quark loop in the light cone gauge.}\label{quark-loop}
\end{figure}
The quark loop in Fig.\ref{quark-loop} can be expressed as

\begin{eqnarray}
\Pi^{\mu\alpha}_{ab}(P)&=&-i\frac{1}{2}N_{f}g^{2}\delta_{ab}\int\frac{d^{4}K}{(2\pi)^{4}}\mathrm{Tr}[\gamma^{\mu}F(K)\gamma^{\alpha}F(K-P)]\;,
\end{eqnarray}
where $N_{f}$ is the active quark flavors and $F(K)$ is the bare quark propagator.

The retarded self energy in the real time formalism \cite{K1965}, \cite{CDT1999} is expressed as
\begin{eqnarray}
&&\Pi_{R}^{\mu\alpha}(P)=\Pi_{11}^{\mu\alpha}(P)+\Pi_{12}^{\mu\alpha}(P)\notag\\
&&=-\frac{i}{2}g^{2}N_{f}\int\frac{d^{4}K}{(2\pi)^{4}}\mathrm{Tr}[\gamma^{\mu}\cancel{K}\gamma^{\alpha}(\cancel{K}-\cancel{P})][\tilde{\Delta}_{11}(K)\tilde{\Delta}
 _{11}(K-P)-\tilde{\Delta}_{12}(K)\tilde{\Delta}_{21}(K-P)]\;,
\end{eqnarray}

where the RTF Green function $\tilde{\Delta}_{ij}(K)$ refers to the component of the propagator in the real time  formalism in Eq.\eqref{RTF-q-propagator}.

Multiply $\Pi^{\mu\alpha}(P)$ by the transverse projection tensor $-\frac{1}{2}\tilde{T}_{P\mu\alpha}$, we can get the transverse self energy $\Pi_{T}(P)$,
\begin{small}
\begin{eqnarray}
\Pi_{T}(P)&=&\frac{i}{4}N_{f}g^{2}\int\frac{\mathrm{d}^{4}K}{(2\pi)^{4}}8\left[K\cdot P\!-\!2\frac{n\cdot K}{n\cdot P}K\cdot P\!+\!\frac{(n\cdot K)^{2}}{(n\cdot P)^{2}}P^{2}\right]\big[\tilde{\Delta}_{11}(K)\tilde{\Delta}_{11}(K\!-\!P)\!-\!\tilde{\Delta}_{12}(K)\tilde{\Delta}_{21}(K\!-\!P)\big]\;.
\end{eqnarray}\end{small}

Using the relation in Eq.\eqref{RTFGreen-functions}, we can obtain
\begin{eqnarray}
&&\tilde{\Delta}_{11}(K)\tilde{\Delta}_{11}(K-P)-\tilde{\Delta}_{12}(K)\tilde{\Delta}_{21}(K-P)\notag\\
&&=\frac{1}{2}\big[\tilde{\Delta}_{S}(K-P)\tilde{\Delta}_{R}(K)+\tilde{\Delta}_{A}(K-P)\tilde{\Delta}_{S}(K)+\tilde{\Delta}_{A}(K-P)\tilde{\Delta}_{A}(K)+\tilde{\Delta}_{R}(K-P)\tilde{\Delta}_{R}(K)\big]\notag\\
&&=\frac{1}{2}\big[\tilde{\Delta}_{S}(K-P)\tilde{\Delta}_{R}(K)+\tilde{\Delta}_{A}(K-P)\tilde{\Delta}_{S}(K)\big]\;,
\end{eqnarray}
where the minus sign in front of the  term $\tilde{\Delta}_{12}(K)\tilde{\Delta}_{21}(K-P)$  comes from the vertex of the type 2 fields \cite{LW1987}. The $k_{0}$ integral of $\tilde{\Delta}_{A}(K-P)\tilde{\Delta}_{A}(K)$ and $\tilde{\Delta}_{R}(K-P)\tilde{\Delta}_{R}(K)$ reduce to zero.

Replace $K$ by $P-K$ in the first term and using $\tilde{\Delta}_{R}(P-K)=\tilde{\Delta}_{A}(K-P)$, and this expression can be simplified further on,
\begin{small}
\begin{eqnarray}
\Pi_{T}(P)&=&\frac{i}{4}N_{f}g^{2}\int\frac{\mathrm{d}^{4}K}{(2\pi)^{4}}8\left[K\!\cdot\! P\!-\!2\frac{n\!\cdot\! K}{n\!\cdot\! P}K\!\cdot\! P\!+\!\frac{(n\!\cdot \!K)^{2}}{(n\!\cdot\! P)^{2}}P^{2}\right]\tilde{\Delta}
 _{S}(K)\tilde{\Delta}
 _{A}(K-P)\notag\\
&=&\frac{i}{4}N_{f}g^{2}\int\frac{\mathrm{d}^{4}K}{(2\pi)^{4}}8\left[K\!\cdot\! P\!-\!2\frac{n\!\cdot\! K}{n\!\cdot\! P}K\!\cdot\! P\!+\!\frac{(n\!\cdot \!K)^{2}}{(n\!\cdot\! P)^{2}}P^{2}\right](-2\pi i)\delta(K^{2})[1\!-\!2f(k_{0})]\frac{1}{(K\!-\!P)^{2}\!-\!isgn(k_{0}\!-\!p_{0})\epsilon}\;.
\end{eqnarray}\end{small}

In the calculation, we use the HTL approximation,i.e. high temperature limit. The internal momentum $K$ is hard, and the external momentum momentum $P$ is soft. The transverse part of the quark loop in the light cone gauge in the HTL approximation from Eq.\eqref{TL} is obtained by
\begin{small}
\begin{eqnarray}
\Pi_{T}(P)&=&\frac{i}{4}N_{f}g^{2}\int\frac{\mathrm{d}^{4}K}{(2\pi)^{4}}8\left[K\cdot P-2\frac{n\cdot K}{n\cdot P}K\cdot P+\frac{(n\cdot K)^{2}}{(n\cdot P)^{2}}P^{2}\right]\tilde{\Delta}
 _{S}(K)\tilde{\Delta}
 _{A}(K\!-\!P)\notag\\
&=&\frac{1}{12}N_{f} g^{2}T^{2}\left[\frac{(p_{0})^{2}}{p^{2}}-\frac{p_{0}P^{2}}{2p^{3}}\ln\frac{p_{0}+p+i\epsilon}{p_{0}-p+i\epsilon}\right]\;.
\end{eqnarray}\end{small}

Similarly, the longitudinal part of the quark loop in the light cone gauge in the HTL approximation is
\begin{small}
\begin{eqnarray}
\frac{1}{n_{P}^{2}}
 \Pi_{L}(P)&\!=\!&\frac{i}{2}g^{2}N_{f}\int\frac{d^{4}K}{(2\pi)^{4}}\left[8\frac{(n\cdot K)^{2}}{(n\cdot P)^{2}}P^{2}\!-\!16\frac{n\cdot K}{n\cdot P}K\cdot P\!+\!8\frac{(K\cdot P)^{2}}{P^{2}}\!-\!4K\cdot P+4K^{2}\right]\tilde{\Delta}
 _{S}(K)\tilde{\Delta}
 _{A}(K\!-\!P)\notag\\
&=&-\frac{1}{6}N_{f}g^{2}T^{2}\left[\frac{P^{2}}{p^{2}}-\frac{p_{0}P^{2}}{2p^{3}}\ln\frac{p_{0}+p+i\epsilon}{p_{0}-p+i\epsilon}\right]\;.
\end{eqnarray}\end{small}

\subsection{The gluon loop and the gluon tadpole of the gluon HTL self energy in the light cone gauge}

\begin{figure}
  \centering
  \includegraphics[width=6cm]{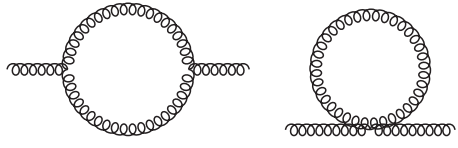}\\
  \caption{The gluon loop and the gluon tadpole in the light cone gauge.}\label{gloun-loop and tadpole}
\end{figure}
The gluon loop and the gluon tadpole   in Fig.\ref{gloun-loop and tadpole} are expressed as
\begin{eqnarray}
\Pi^{\mu\alpha}_{ab}(P)\!&=&\!\frac{i}{2}\int\frac{d^{4}K}{(2\pi)^{4}}V^{\mu\nu\rho}(P,\!-\!K,K\!-\!P)id_{\nu\beta}(K)V^{\beta\gamma\alpha}(K,P\!-\!K,\!-\!P)id_{\rho\gamma}(K\!-\!P)G(K)G(K\!-\!P)\notag\\
&+&\frac{i}{2}\int\frac{d^{4}K}{(2\pi)^{4}}id_{\rho\sigma}(K)G(K)\delta^{cd}V^{\mu\alpha\rho\sigma}_{abcd}\;.
\end{eqnarray}
where $d_{\nu\beta}(K)G(K)$, $d^{\rho\gamma}(K-P)G(K-P)$ and $d_{\rho\sigma}(K)G(K)$ are all the bare gluon propagators.

Below are the tensor of the bare gluon propagator in the light cone gauge
\begin{equation}
d_{\nu\beta}(K)\!=\!-g_{\nu\beta}+\frac{n_{\nu}K_{\beta}+n_{\beta}K_{\nu}}{n\cdot K}\;.\\
\end{equation}

The three gluon vertexes  are
\begin{eqnarray}\label{three-g-vertexes}
V^{\mu\nu\rho}(P,-K,K-P)&=&gf^{acd}\big[g^{\mu\nu}(P+K)^{\rho}+g^{\nu\rho}(-2K+P)^{\mu}+g^{\mu\rho}(K-2P)^{\nu}\big]\notag\;,\\
V^{\beta\gamma\alpha}(K,P-K,-P)&=&gf^{cdb}\big[g^{\beta\gamma}(2K-P)^{\alpha}+g^{\gamma\alpha}(2P-K)^{\beta}+g^{\beta\alpha}(-P-K)^{\gamma}\big]\;.
\end{eqnarray}
The four gluon vertex is
\begin{equation}
V^{\mu\alpha\rho\sigma}_{abcd}\!=\!-ig^{2}[f^{abe}f^{cde}(g^{\mu\rho}g^{\alpha\sigma}\!-\!g^{\mu\sigma}g^{\alpha\rho})\!+\!f^{ace}f^{bde}(g^{\mu\alpha}g^{\rho\sigma}\!-\!g^{\mu\sigma}g^{\alpha\rho})\!+\!f^{ade}f^{bce}(g^{\mu\alpha}g^{\rho\sigma}\!-\!g^{\mu\rho}g^{\alpha\sigma})]\;.
\end{equation}

The transverse part of the gluon self energy in the HTL approximation from Eq.\eqref{TL} is

\begin{small}
\begin{eqnarray}\label{gloopTransverse}
\Pi_{T}(P)&\!=\!&\frac{i}{4}C_{A}g^{2}\int\frac{d^{4}K}{(2\pi)^{4}}\Big[-8K\!\cdot\! P\!-\!12P^{2}\!+\!16\frac{n\!\cdot\! K}{n\!\cdot\!P}K\!\cdot\! P\!-\!8\frac{(n\!\cdot\! K)^{2}}{(n\!\cdot\!P)^{2}}P^{2}\notag\\
&&+8\frac{n\!\cdot\! P}{n\!\cdot\!(K\!-\!P)}(K^{2}\!-\!P^{2})\!-\!8\frac{n\!\cdot\! P}{n\!\cdot\!K}(K^{2}\!-\!2K\!\cdot\! P)\Big]G(K)G(K\!-\!P)\notag\\
&&=\frac{1}{6}C_{A}g^{2}T^{2}\left[\frac{(p_{0})^{2}}{p^{2}}-\frac{p_{0}P^{2}}{2p^{3}}\ln\frac{p_{0}+p+i\epsilon}{p_{0}-p-i\epsilon}\right]\notag\\
&&+\frac{i}{4}C_{A}g^{2}\int\frac{d^{4}K}{(2\pi)^{4}}\left[8\frac{n\!\cdot\! P}{n\!\cdot\!(K\!-\!P)}(K^{2}\!-\!P^{2})\!-\!8\frac{n\!\cdot\! P}{n\!\cdot\!K}(K^{2}\!-\!2K\!\cdot\! P)\right]G(K)G(K\!-\!P)\;.
\end{eqnarray}
\end{small}
The last two light cone terms in the forth line come from the the tensors of the  gluon propagators $(n^{\nu}K^{\beta}+n^{\beta}K^{\nu})/(n\cdot K)$ and $[n^{\rho}(K-P)^{\gamma}+n^{\gamma}(K-P)^{\rho}]/((n\cdot (K-P))$.
Replace $K$ by $P-K$ in the first light cone  term, and
the sum of the two light cone terms is
\begin{small}
\begin{eqnarray}\label{light-cone-termT}
-\frac{i}{4}C_{A}g^{2}\int\frac{\mathrm{d}^{4}K}{(2\pi)^{4}}16\frac{n\!\cdot\! P}{n\!\cdot\!K}\frac{(K^{2}\!-\!2K\!\cdot\! P)}{K^{2}(K\!-\!P)^{2}}\;,
\end{eqnarray}\end{small}
At zero temperature, the divergence of the integral calculation of one loop diagram in Fig.\ref{gloun-loop and tadpole} has been renormalized successfully.
Here  we only consider the contribution at finite temperature.
Use the ML prescription of $1/(n\cdot K)$ in Eq.\eqref{MLprescription}. By calculating with the contour integral, we find there is no divergence in the HTL approximation,  and the power of the  part  is  $g^{3}T^{2}$ order, which can be ignored in the result. The proof is in the Appendix.

Similarly, the longitudinal part of the gluon self energy  in the HTL approximation is obtained by

\begin{small}
\begin{eqnarray}\label{gloopLongitudinal}
\frac{1}{n_{P}^{2}}\Pi_{L}(P)&\!=\!&\frac{i}{2}C_{A}g^{2}\int\frac{d^{4}K}{(2\pi)^{4}}[4K^{2}\!-\!4K\!\cdot\! P\!+\!2P^{2}\!+\!8\frac{(K\!\cdot\! P)^{2}}{P^{2}}\!-\!16\frac{n\!\cdot\! K}{n\!\cdot\! P}K\!\cdot\! P\!+\!8\frac{(n\!\cdot\! K)^{2}}{(n\!\cdot\! P)^{2}}P^{2}\notag\\
&+&\frac{n\!\cdot\! P}{n\!\cdot\! (K-P)}\frac{2K^{2}(K\!-\!P)^{2}}{P^{2}}\!-\!\frac{n\!\cdot\! P}{n\!\cdot\!K}\frac{2K^{2}(K\!-\!P)^{2}}{P^{2}}]G(K)G(K\!-\!P)\notag\\
&\!=\!&-\frac{1}{3}C_{A}g^{2}T^{2}\left[\frac{P^{2}}{p^{2}}-\frac{p_{0}P^{2}}{2p^{3}}\ln\frac{p_{0}+p+i\epsilon}{p_{0}-p+i\epsilon}\right]+\frac{i}{2}C_{A}g^{2}\int\frac{d^{4}K}{(2\pi)^{4}}\frac{1}{P^{2}}\left[2\frac{n\!\cdot\! P}{n\!\cdot\! (K-P)}\!-\!2\frac{n\!\cdot\! P}{n\!\cdot\!K}\right]\;.
\end{eqnarray}
\end{small}
There are  two light cone terms in the third line.
By calculating, the integral with the light cone terms  is zero.
\begin{equation}
\frac{i}{2}C_{A}g^{2}\int\frac{d^{4}K}{(2\pi)^{4}}\frac{1}{P^{2}}\left[2\frac{n\!\cdot\! P}{n\!\cdot\! (K-P)}\!-\!2\frac{n\!\cdot\! P}{n\!\cdot\!K}\right]=0\;.
\end{equation}

During the calculation, we set the momentum $\overrightarrow{p}$ is on the positive direction of $z$ axis.
$\overrightarrow{n}$ in Eq.\eqref{lcg-axial-vector} is on the negative direction of $z$ axis.
The angle between $\overrightarrow{k}$ and $\overrightarrow{p}$ is $\theta$.
Here we use the Kelydsh representation in the real time formalism  and consider the $T>0$ contribution.
In the HTL approximation, the internal momentum $K$ is soft,
and the external momentum $P$ is hard.

\subsection{The  transverse and longitudinal  parts of the gluon HTL self energy}

\indent Adding up the results from the longitudinal and transverse parts of the quark loop in Fig.\ref{quark-loop},
the gluon loop and the gluon tadpole  in Fig.\ref{gloun-loop and tadpole}   in the HTL approximation, we obtain the following expression,
\begin{small}
\begin{eqnarray}\label{TLselfenergy}
\Pi_{T}(P)&\!=\!&\frac{1}{6}(C_{A}+\frac{1}{2}N_{f})g^{2}T^{2}\left[\frac{(p_{0})^{2}}{p^{2}}-\frac{p_{0}P^{2}}{2p^{3}}\ln\frac{p_{0}+p+i\epsilon}{p_{0}-p+i\epsilon}\right]\notag\;,\\
\frac{1}{n_{P}^{2}}
 \Pi_{L}(P)&\!=\!&-\frac{1}{3}(C_{A}+\frac{1}{2}N_{f})g^{2}T^{2}\left[\frac{P^{2}}{p^{2}}-\frac{p_{0}P^{2}}{2p^{3}}\ln\frac{p_{0}+p+i\epsilon}{p_{0}-p+i\epsilon}\right]\;.
\end{eqnarray}
\end{small}

The factor $\frac{1}{2}$ in the coefficient $(C_{A}+\frac{1}{2}N_{f})$ stems from that these integrals $\int_{0}^{\infty} k n(k)\mathrm{d}k $ and $\int_{0}^{\infty} k f(k)\mathrm{d}k$ which have different distribution functions.
We  find we get the same result of the transverse and longitudinal HTL gluon self energy in the light cone gauge and the Coulomb gauge,
although these two gauges have different projection tensors.

In the static limit $p_{0}\rightarrow 0$, the longitudinal and transverse parts of  gluon HTL self energy in the light cone gauge reduce to
\begin{eqnarray}
\lim_{p_{0}\to 0} \frac{1}{n_{P}^{2}}\Pi_{L}(P) &=& \frac{1}{3}g^{2}T^{2}  (C_{A}+\frac{1}{2} N_{f})=m_{D}^{2}\;,\notag\\
\lim_{p_{0}\to 0} \Pi_{L}(P) &=& \frac{1}{6}g^{2}T^{2}  (C_{A}+\frac{1}{2} N_{f})=\frac{1}{2}m_{D}^{2}\;,\notag\\
\lim_{p_{0}\to 0} \Pi_{T}(P)&=& 0\;.
\end{eqnarray}

In the static limit  $p_{0}\rightarrow 0$, the longitudinal and transverse  parts of the  gluon HTL self energy in the Coulomb gauge reduces to
\begin{eqnarray}
\lim_{p_{0}\to 0}\Pi_{L}(P)&=&\frac{1}{3}(C_{A}+\frac{1}{2} N_{f})g^{2} T^{2}=m_{D}^{2}\notag\;,\\
\lim_{p_{0}\to 0}\Pi_{T}(P)&=&0\;.
\end{eqnarray}

We think the axial vector $n_{l}^{\mu}=(\frac{\sqrt{2}}{2},0,0,-\frac{\sqrt{2}}{2})$ in the light cone gauge is rotated with respect to the axial vector $n_{c}^{\mu}=(1,0,0,0)$ in the Coulomb gauge,
so that it gives rise to some changes.

In the static limit, $n_{\mu}n_{\nu}\Pi^{\mu\nu}(P)$ of the gluon HTL self energy in the Coulomb gauge in Eq.\eqref{selfenergyCoulomb} is
\begin{eqnarray}
\lim_{p_{0}\to 0}n_{\mu}n_{\nu}\Pi^{\mu\nu}(P)&=&\lim_{p_{0}\to 0}\Pi^{00}(P)=-\lim_{p_{0}\to 0}\frac{L_{P}^{00}}{n_{P}^{2}}\Pi_{L}(P)=-m_{D}^{2}\;,
\end{eqnarray}
where the axial vector $n_{c}^{\mu}=(1,0,0,0)$ in Eq.\eqref{axial-vector}.

Because the axial vector $n_{l}^{\mu}$ in the light cone gauge is different from $n_{c}^{\mu}$ in the Coulomb gauge, we compare $\Pi^{00}(P)$ in the light cone gauge with $\Pi^{00}(P)$ in the Coulomb gauge.

In the static limit, $\Pi^{00}(P)$ of the gluon HTL self energy in the light cone gauge in Eq.\eqref{selfenergylgc} is
\begin{eqnarray}
\lim_{p_{0}\to 0} \Pi^{00}(P)=\lim_{p_{0}\to 0}-\tilde{L}_{P}^{00}\frac{1}{n_{P}^{2}}\Pi_{L}(P)=-\frac{g^{2}T^{2}}3 (C_{A}+\frac{1}{2} N_{f})=-m_{D}^{2}\;,
\end{eqnarray}
where the axial vector $n_{l}^{\mu}=(\frac{\sqrt{2}}{2},0,0,-\frac{\sqrt{2}}{2})$ in Eq.\eqref{lcg-axial-vector}. The external momentum $P^{\mu}=(p_{0},0,0,p_{3})$, $\overrightarrow{p}
 $ is on the $z$ axis.
The result is the same as that in the Coulomb gauge.

\subsection{The HTL resumed gluon propagator in the light cone gauge and its spectral function }

The HTL resumed gluon propagator in the  light cone gauge is
\begin{eqnarray}\label{HTL-resummed-g-propagator}
&&\Delta^{\mu\nu}(P)=\frac{\tilde{T}_{P}^{\mu\nu}}{-P^{2}+\Pi_{T}(P)}+\frac{-\frac{n^{\mu}n^{\nu}P^{2}}{(n\cdot P)^{2}}}{-P^{2}+\frac{1}{n_{P}^{2}}\Pi_{L}(P)}\notag\\
&&=\frac{-\tilde{T}_{P}^{\mu\nu}}{P^{2}-\frac{1}{2}m_{D}^{2}\left[\frac{(p_{0})^{2}}{p^{2}}-\frac{p_{0}P^{2}}{2p^{3}}\ln\frac{p_{0}+p+i\epsilon}{p_{0}-p+i\epsilon}\right]}+\frac{\frac{n^{\mu}n^{\nu}P^{2}}{(n\cdot P)^{2}}}{P^{2}+m_{D}^{2}\left[\frac{P^{2}}{p^{2}}-\frac{p_{0}P^{2}}{2p^{3}}\ln\frac{p_{0}+p+i\epsilon}{p_{0}-p+i\epsilon}\right]}\;.
\end{eqnarray}

The transverse and longitudinal spectral functions are expressed as
\begin{eqnarray}
\rho_{T}(P)&=&2\pi Z_{T} \mathrm{sgn}(p_{0})\big[\delta(p_{0}-w_{T})+\delta(p_{0}+w_{T})\big]+\beta_{T}(P)\;,\notag\\
\rho_{L}(P)&=&2\pi Z_{L}\mathrm{sgn}(p_{0})\big[\delta(p_{0}-w_{L})+\delta(p_{0}+w_{L})\big]+\beta_{L}(P)\;.
\end{eqnarray}
The spectral function $\rho_{T/L}(P)$ is made up of the pole term  and the cut term   $\beta_{T/L}(P)$.

For $P^{2}$ space-like,i.e $p_{0}^{2}< p^{2}$, the function $\ln\frac{p_{0}+p+i\epsilon}{p_{0}-p+i\epsilon}$ generates the imaginary part,
\begin{equation}\label{imaginarypart}
\ln\frac{p_{0}+p\pm i\epsilon}{p_{0}-p\pm i\epsilon}=\ln|\frac{p_{0}+p}{p_{0}-p}|\mp i\pi\theta(p^{2}-p_{0}^{2})\;.
\end{equation}

So the cut terms of the transverse and longitudinal part of the HTL resumed gluon propagator in the light cone gauge are obtained by
\begin{small}
\begin{eqnarray}
&&\beta_{T}(P)\!=\!\frac{\frac{1}{2}\pi m_{D}^{2}\frac{p_{0}P^{2}}{p^{3}}\theta(p^{2}\!-\!p_{0}^{2})}{\bigg[P^{2}\!-\!\frac{1}{2}m_{D}^{2}\Big[\frac{(p_{0})^{2}}{p^{2}}\!-\!\frac{p_{0}P^{2}}{2p^{3}}\ln|\frac{p_{0}\!+\!p}{p_{0}\!-\!p}|\Big]\bigg]^{2}\!+\!\frac{1}{16}m_{D}^{4}\pi^{2}\frac{(p_{0})^{2}P^{2}P^{2}}{p^{6}}}\notag\;,\\
&&\beta_{L}(P)\!=\!\frac{-\pi m_{D}^{2}\frac{p_{0}}{p}\theta(p^{2}\!-\!p_{0}^{2})}{\bigg[ p^{2}\!+\!m_{D}^{2}\Big[1\!-\!\frac{p_{0}}{2p}\ln|\frac{p_{0}\!+\!p}{p_{0}\!-\!p}|\Big]\bigg]^{2}\!+\!\frac{1}{4}m_{D}^{4}\pi^{2}\frac{(p_{0})^{2}}{p^{2}}}\;,
\end{eqnarray}\end{small}
where  $\theta(p^{2}-p_{0}^{2})$ is the step function, and the Debye screening mass $m_{D}^{2}=\frac{1}{3}(C_{A}\!+\!\frac{1}{2}N_{f})g^{2}T^{2}$.

Via the expression of the HTL resumed gluon propagator in the light cone gauge,
the transverse dispersion relation is given by
\begin{equation}
\omega_{T}^{2}-p^{2}-m_{D}^{2}\Big[\frac{\omega_{T}^{2}}{p^{2}}-\frac{\omega_{T}(\omega_{T}^{2}-p^{2})}{2p^{3}}\ln\frac{\omega_{T}+p}{\omega_{T}-p}\Big]=0\;,
\end{equation}

where $\omega_{T}$ is the solution of the above transverse dispersion relation.

The longitudinal dispersion relation is given by
\begin{equation}
p^{2}+m_{D}^{2}\Big[1-\frac{\omega_{L}}{2p}\ln\frac{\omega_{L}+p}{\omega_{L}-p}\Big]=0\;,
\end{equation}

where $\omega_{L}$ is the solution of the above longitudinal dispersion relation.

Obviously  the  transverse and longitudinal parts have the same dispersion relation as that of the HTL resumed gluon propagator  in the Coulomb gauge. However, this two kinds of gauge have different expression of  the transverse and longitudinal projection tensor, and then have different  expression of the HTL resumed gluon propagator. Due to the same dispersion relation, for more analyses you can refer to \cite{B1996}.

 The residue for the transverse part is

\begin{eqnarray}
&&Z_{T}=-\bigg(\Big[\frac{\partial(P^{2}-\Pi_{T})}{\partial p_{0}}\Big]_{p_{0}=\omega_{T}(p)}\bigg)^{-1}\notag\\
&&=\frac{\omega_{T}(\omega_{T}^{2}-p^{2})}{m_{D}^{2}\omega_{T}^{2}-(\omega_{T}^{2}-p^{2})^{2}}\;.
\end{eqnarray}

The residue for the longitudinal part is
\begin{eqnarray}
&&Z_{L}=-\bigg(\Big[\frac{\partial\frac{p^{2}}{P^{2}}(P^{2}-\frac{1}{n_{p}^{2}}\Pi_{L})}{\partial p_{0}}\Big]_{p_{0}=\omega_{L}(p)}\bigg)^{-1}\notag\\
&&=\frac{\omega_{L}(\omega_{L}^{2}-p^{2})}{p^{2}(p^{2}-m_{D}^{2}-\omega_{L}^{2})}\;.
\end{eqnarray}
About the transverse and longitudinal residues you can find more discussion in \cite{B1996} too.

Below is the proof to tell the reason why we can get the same  dispersion relation.
From the famous Ward identity,
\begin{equation}
P^{\mu}\Pi_{\mu\nu}(P)=p_{0}\Pi_{0\nu}-p_{3}\Pi_{_{3\nu}}=0\;,
\end{equation}

we can get the below relation,
\begin{equation}
\Pi_{_{3\nu}}=\frac{p_{0}}{p_{3}}\Pi_{0\nu}\;,
\end{equation}
where the external momentum $P^{\mu}=(p_{0},0,0,p_{3})$,  $p_{3}>0$, and $\overrightarrow{p}$  is on the positive direction of z axis.

Using the above relation, the longitudinal part of the gluon HTL self energy can become
\begin{eqnarray}
&&\tilde{L}_{P}^{\mu\nu}\Pi_{\mu\nu}(P)=-\frac{n^{\mu}n^{\nu}P^{2}}{(n\cdot P)^{2}}\Pi_{\mu\nu}(P)\notag\\
&&=-\frac{n^{\nu}P^{2}}{(n\cdot P)^{2}}\frac{\sqrt{2}}{2}(\Pi_{0\nu}+\Pi_{3\nu})\notag\\
&&=-\frac{P^{2}}{(n\cdot P)^{2}}\frac{\sqrt{2}(p_{0}+p_{3})}{2p_{3}}n^{\nu}\Pi_{0\nu}\notag\\
&&=-\frac{P^{2}}{(n\cdot P)^{2}}\frac{(p_{0}+p_{3})^{2}}{2(p_{3})^{2}}\Pi^{00}(P)\notag\\
&&=-\frac{P^{2}}{(p_{3})^{2}}\Pi^{00}(P)\;,
\end{eqnarray}
where $L_{P}^{\mu\nu}$ is the longitudinal projection tensor in the light cone gauge in Eq.\eqref{longitudinal},
$\Pi^{00}(P)$ is the longitudinal part of the  gluon HTL self energy in the Coulomb gauge in Eq.\eqref{TL-parts},
and the axial vector in the light cone gauge $n_{l}^{\mu}=(\frac{\sqrt{2}}{2},0,0,-\frac{\sqrt{2}}{2})$.
The result tells us the longitudinal part of the gluon HTL self energy in the light cone gauge is the same as  that in the Coulomb gauge.

Similarly, the transverse part of the gluon HTL  self energy can become
\begin{eqnarray}
&&\frac{1}{2}\tilde{T}_{P}^{\mu\nu}\Pi_{\mu\nu}(P)=\frac{1}{2}\Big[g^{\mu\nu}+\frac{n^{\mu}n^{\nu}P^{2}}{(n\cdot P)^{2}}\Big]\Pi_{\mu\nu}(P)\notag\\
&&=-\frac{1}{2}m_{D}^{2}+\frac{1}{2}\frac{P^{2}}{(p_{3})^{2}}\Pi^{00}(P)\;.
\end{eqnarray}
where $\tilde{T}_{P}^{\mu\nu}$ is the transverse projection tensor in the light cone gauge in Eq.\eqref{transverse}. So the transverse part of the gluon HTL self energy in the light cone gauge is the same as that in the Coulomb gauge.

In the proof, we  use the famous Ward identity.
 $\overrightarrow{p}$ is on the positive direction of the z axis, and
$\overrightarrow{n}$ is on the negative direction of z axis.
We can get the same transverse and longitudinal parts of the gluon HTL self energy in the two gauges,
and finally get the same dispersion relation  of the transverse and longitudinal parts.

\section{The HTL resumed quark propagator in the light cone gauge}

\begin{figure}
  \centering
  \includegraphics[width=4cm]{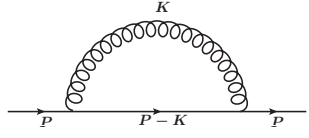}\\
  \caption{The quark self energy in the light cone gauge.}\label{quarkselfenergy}
\end{figure}

The quark self energy in Fig.\ref{quarkselfenergy} can be expressed as

\begin{equation}
\Sigma(P)=iC_{F}g^{2}\int\frac{\mathrm{d}^{4}K}{(2\pi)^{4}}\gamma^{\mu}F(P-K)\gamma^{\nu}d_{\mu\nu}(K)G(K)\;,
\end{equation}
where $C_{F}=\frac{4}{3}$ is the color factor, $F(P-K)$ is the bare quark propagator, and $d_{\mu\nu}(K)G(K)$ is the bare gluon propagator.

The retarded quark self energy in the real time formalism is expressed as
\begin{eqnarray}
&&\Sigma_{R}(P)=\Sigma_{11}(P)+\Sigma_{12}(P)\notag\\
&&=iC_{F}g^{2}\int\frac{\mathrm{d}^{4}K}{(2\pi)^{4}}\gamma^{\mu}(\cancel{P}-\cancel{K})\gamma^{\nu}d_{\mu\nu}(K)[\tilde{\Delta}_{11}(P-K)\Delta_{11}(K)-\tilde{\Delta}_{12}(P-K)\Delta_{12}(K)]\;.
\end{eqnarray}

Using the relation in Eq.\eqref{RTFGreen-functions}, we get
\begin{eqnarray}
&&\tilde{\Delta}_{11}(P-K)\Delta_{11}(K)-\tilde{\Delta}_{12}(P-K)\Delta_{12}(K)\notag\\
&&=\frac{1}{2}[\tilde{\Delta}_{R}\Delta_{S}+\tilde{\Delta}_{R}\Delta_{A}+\tilde{\Delta}_{S}\Delta_{R}+\tilde{\Delta}_{A}\Delta_{R}]\notag\\
&&=\frac{1}{2}[\tilde{\Delta}_{R}\Delta_{S}+\tilde{\Delta}_{S}\Delta_{R}]\;,
\end{eqnarray}
where $\tilde{\Delta}$ and $\Delta$ respectively represent  the Green functions of quark and gluon.The terms $\tilde{\Delta}_{R}\Delta_{A}$ and $\tilde{\Delta}_{A}\Delta_{R}$  are both zero temperature parts, and we neglect them here.

The quark self energy  is decomposed into two parts,
\begin{eqnarray}
&&\Sigma_{R}(P)=-a(p_{0},p)\cancel{P}-b(p_{0},p)\gamma^{0}\;,\notag\\
&&a(p_{0},p)=\frac{1}{4p^{2}}\big[\mathrm{Tr}(\cancel{P}\Sigma_{R})-p_{0}\mathrm{Tr}[\gamma_{0}\Sigma_{R}]\big]\;,\notag\\
&&b(p_{0},p)=\frac{1}{4p^{2}}\big[P^{2}\mathrm{Tr}[\gamma_{0}\Sigma_{R}]-\gamma_{0}\mathrm{Tr}(\cancel{P}\Sigma_{R})\big]\;.
\end{eqnarray}

With the above  relation, we can do the following calculation,
\begin{small}
\begin{eqnarray}
\mathrm{Tr}[\cancel{P}\Sigma_{R}(P)]=iC_{F}g^{2}\int\frac{\mathrm{d}^{4}K}{(2\pi)^{4}}\big[-8K\cdot P+\frac{n\cdot P}{n\cdot K}(-8K^{2}+16K\cdot P)\big]\frac{1}{2}\big[\tilde{\Delta}_{R}(P-K)\Delta_{S}(K)+\tilde{\Delta}_{S}(P-K)\Delta_{R}(K)\big]\;.
\end{eqnarray}\end{small}
Replace $K$ by $P-K$ in this term $\tilde{\Delta}_{S}(P-K)\Delta_{R}(K)$, this expression becomes

\begin{small}
\begin{eqnarray}
&&\mathrm{Tr}[\cancel{P}\Sigma_{R}(P)]=iC_{F}g^{2}\int\frac{\mathrm{d}^{4}K}{(2\pi)^{4}}\bigg[\Big[8(K\cdot P-P^{2})+\frac{8n\cdot P}{n\cdot(K-P)}(K^{2}-P^{2})
 \Big]\frac{1}{2}\tilde{\Delta}_{S}(K)\Delta_{R}(P-K)\notag\\
&&+\Big[-8K\cdot P+\frac{n\cdot P}{n\cdot K}(-8K^{2}+16K\cdot P)\Big]\frac{1}{2}\tilde{\Delta}_{R}(P-K)\Delta_{S}(K)\bigg]\notag\\
&&=iC_{F}g^{2}\int\frac{\mathrm{d}^{4}K}{(2\pi)^{4}}\bigg[\Big[4(K\cdot P-P^{2})\tilde{\Delta}_{S}(K)\Delta_{R}(P-K)-4K\cdot P\tilde{\Delta}_{R}(P-K)\Delta_{S}(K)\Big]\notag\\
&&+\Big[\frac{4n\cdot P}{n\cdot(K-P)}(K^{2}-P^{2})
\tilde{\Delta}_{S}(K)\Delta_{R}(P-K)+\frac{n\cdot P}{n\cdot K}(-4K^{2}+8K\cdot P)\tilde{\Delta}_{R}(P-K)\Delta_{S}(K)\Big]\bigg]\notag\\
&&=4m_{F}^{2}\;,
\end{eqnarray}\end{small}

In the HTL approximation, we can prove there is no spurious divergence from the light cone terms in the forth line,
and these finite terms are power suppressed than the covariant terms, so we ignore these light cone terms.

In the same way, we can get
\begin{small}
\begin{eqnarray}
&&\mathrm{Tr}[\gamma_{0}\Sigma_{R}(P)]=C_{F}g^{2}\int\frac{\mathrm{d}^{4}K}{(2\pi)^{4}}8[-k_{0}+k_{0}\frac{n\cdot P}{n\cdot K}-\frac{n_{0}}{n\cdot K}(K^{2}-K\cdot P)]\frac{1}{2}\big[\tilde{\Delta}_{R}(P-K)\Delta_{S}(K)+\tilde{\Delta}_{S}(P-K)\Delta_{R}(K)\big]\notag\\
&&=C_{F}g^{2}\int\frac{\mathrm{d}^{4}K}{(2\pi)^{4}}\bigg[4(k_{0}-p_{0})\tilde{\Delta}_{S}(K)\Delta_{R}(P-K)-4k_{0}\tilde{\Delta}_{R}(P-K)\Delta_{S}(K)\notag\\
&&+4\Big[(k_{0}-p_{0})\frac{n\cdot P}{n\cdot(K-P)}+\frac{n_{0}}{n\cdot(K-P)}(K^{2}-K\cdot P)\Big]\tilde{\Delta}_{S}(K)\Delta_{R}(P-K)\notag\\
&&+4\Big[k_{0}\frac{n\cdot P}{n\cdot K}-\frac{n_{0}}{n\cdot K}(K^{2}-K\cdot P)\Big]\tilde{\Delta}_{R}(P-K)\Delta_{S}(K)\bigg]\notag\\
&&=2m_{F}^{2}\frac{1}{p}\ln\frac{p_{0}+p+i\epsilon}{p_{0}-p+i\epsilon}\;.
\end{eqnarray}\end{small}

In the HTL approximation, it can be  proved that there is no spurious divergence from the light cone terms in the third and  forth lines,
and these finite terms are power suppressed than the covariant terms, so we ignore these light cone terms too.

So the result shows the quark HTL self energy   is the same  as  that of covariant gauge \cite{W1982}.
With the quark HTL self energy, we can derive the same quark resumed propagator.

The HTL resumed quark propagator  is
\begin{eqnarray}
S^{*}(P)&=&\frac{1}{D_{+}(P)}\frac{\gamma_{0}-\hat{\overrightarrow{p}}\cdot\overrightarrow{\gamma}}{2}+\frac{1}{D_{-}(P)}\frac{\gamma_{0}+\hat{\overrightarrow{p}}\cdot\overrightarrow{\gamma}}{2}\notag\;,\\
D_{\pm}(P)&=&-p_{0}\pm p+\frac{m_{F}^{2}}{p}
 \Big[\frac{1}{2}\ln\frac{p_{0}+p+i\epsilon}{p_{0}-p+i\epsilon}\mp(\frac{p_{0}}{2p}\ln\frac{p_{0}+p+i\epsilon}{p_{0}-p-i\epsilon}-1)\Big]\;,
\end{eqnarray}
where the effective  quark mass $m_{F}=g^{2}T^{2}/6$ in QCD, and the notation $\hat{\overrightarrow{p}}=\frac{\overrightarrow{p}}{|\overrightarrow{p}|}$.

\section{The damping  rates of  hard quark and gluon in the light cone gauge}
The damping rates of the heavy fermion have been done \cite{T2000},\cite{P1989}.
In a similar way, we use the HTL resumed gluon propagator in the light cone gauge to calculate the damping rates of the hard quark and gluon.
By above analyses, we know the HTL resumed gluon propagator in the Coulomb gauge and the light cone gauge have the same denominator and different projection tensors in the nominator.
But we can prove in general case, using the HTL resumed gluon propagator in the two gauges get the same result of the damping rates of the hard quark and gluon.

\begin{figure}
  \centering
  \includegraphics[width=6cm]{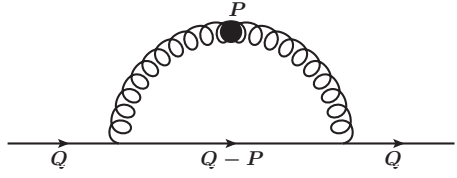}\\
  \caption{The contribution to the damping rate of the  hard quark in the light cone gauge.}\label{quarkdamping}
\end{figure}

The quark self energy in Fig.\ref{quarkdamping} is expressed as

\begin{equation}
\Sigma(Q)=ig^{2}C_{F}\int\frac{\mathrm{d}^{4}P}{(2\pi)^{4}}[\gamma_{\mu}F(Q-P)\gamma_{\nu}G_{\mu\nu}(Q)]\;,
\end{equation}
where $G_{\mu\nu}(Q)$ is the HTL resumed gluon propagator in the light cone gauge, and  $F(Q-P)$ is the bare quark propagator.

In the real time formalism, the retarded quark self energy  is expressed as
\begin{eqnarray}
&&\Sigma_{R}(Q)=\Sigma_{11}(Q)+\Sigma_{12}(Q)\notag\\
&&=ig^{2}C_{F}\int\frac{\mathrm{d}^{4}P}{(2\pi)^{4}}[\gamma_{\mu}(\cancel{Q}-\cancel{P})\gamma_{\nu}][\tilde{\Delta}_{11}(Q-P)\Delta_{11}^{\mu\nu}(P)-\tilde{\Delta}_{12}(Q-P)\Delta_{12}^{\mu\nu}(P)]\;.
\end{eqnarray}

Using the relation of the Keldysh representation in Eq.\eqref{RTFGreen-functions}, we have
\begin{equation}
\tilde{\Delta}_{11}(Q-P)\Delta_{11}^{\mu\nu}(P)-\tilde{\Delta}_{12}(Q-P)\Delta_{12}^{\mu\nu}(P)=\frac{1}{2}[\tilde{\Delta}_{R}\Delta_{S}^{\mu\nu}+\tilde{\Delta}_{R}\Delta_{A}^{\mu\nu}+\tilde{\Delta}_{S}\Delta_{R}^{\mu\nu}+\tilde{\Delta}_{A}\Delta_{R}^{\mu\nu}]\;.
\end{equation}

By power counting the leading contribution  at finite temperature comes from the first term in the bracket $\tilde{\Delta}_{R}\Delta_{S}^{\mu\nu}$, which is $O(1/g^{3})$.
$\tilde{\Delta}_{R}\Delta_{A}^{\mu\nu}$ and $\tilde{\Delta}_{A}\Delta_{R}^{\mu\nu}$ are both $O(1/g^{2})$,
so we ignore them here. From the below equations,  $\tilde{\Delta}_{S}\Delta_{R}^{\mu\nu}$ with the Fermi-Dirac distribution function is $O(g)$ power suppressed than  $\tilde{\Delta}_{R}\Delta_{S}^{\mu\nu}$ with the Bose-Einstein distribution function.

The internal momentum $P$ is soft, $p_{0}\sim gT$. The Bose-Einstein distribution function $n_{B}(p_{0})$ in the term $\tilde{\Delta}_{R}\Delta_{S}^{\mu\nu}$ is $O(1/g)$,
\begin{equation}
\frac{1}{e^{\frac{|p_{0}|}{T}}-1}\sim\frac{T}{|p_{0}|}\propto \frac{1}{g}\;.
\end{equation}

However, the Fermi-Dirac distribution function $f(p_{0}-q_{0})$ in the term $\tilde{\Delta}_{S}\Delta_{R}^{\mu\nu}$ is $O(1)$. The external momentum $Q$ is hard, $q_{0}\sim T$,
\begin{equation}
\frac{1}{e^{\frac{|q_{0}-p_{0}|}{T}}+1}\sim O(1)\;.
\end{equation}

The  symmetric propagator of the soft gluon in the light cone gauge is
\begin{equation}
\Delta_{S}^{\mu\nu}(P)=-2\pi i\Big[-\tilde{T}_{P}^{\mu\nu}\rho_{T}(P)+\frac{n^{\mu}n^{\nu}P^{2}}{(n\cdot P)^{2}}\frac{p^{2}}{P^{2}}\rho_{L}(P)\Big][1+2n_{B}(p_{0})]\;,
\end{equation}
where we only consider the contribution at finite temperature.

Using the below equation, we can calculate the imaginary part of  the quark self energy,
\begin{equation}
\mathrm{Im}\tilde{\Delta}(Q-P)=\mathrm{Im}\Big[\frac{1}{(Q-P)^{2}+i\mathrm{sgn}(q_{0}-p_{0})\epsilon}\Big]=-\pi\mathrm{sgn}(q_{0}-p_{0})\delta[(Q-P)^{2}]\;.
\end{equation}
In the integral, we use the $\delta$ function to integrate out $\cos\theta$,
\begin{equation}
\delta[(Q-P)^{2}]=\frac{1}{2pq}\delta[\cos\theta-\frac{p_{0}}{p}+\frac{P^{2}}{2pq}]\approx\frac{1}{2pq}\delta[\cos\theta-\frac{p_{0}}{p}]\;,
\end{equation}

where the term $\frac{P^{2}}{2pq}\sim g$,  which we can ignore.

The transverse projection tensor in the light cone gauge is
\begin{eqnarray}
\tilde{T}_{P}^{\mu\nu}=g^{\mu\nu}-\frac{n^{\mu}P^{\nu}+n^{\nu}P^{\mu}}{n\cdot P}+\frac{n^{\mu}n^{\nu}P^{2}}{(n\cdot P)^{2}}\;,
\end{eqnarray}
where the internal soft momentum $P^{\mu}=( p_{0},0,0,p_{3})$.

By calculating we can find the relation $\tilde{T}_{P}^{0\nu}=0,\tilde{T}_{P}^{3\nu}=0$, so we have
\begin{equation}
\tilde{T}_{P}^{ij}=-\delta^{ij}, \hspace{1cm} (i,j=1,2).
\end{equation}

The longitudinal and transverse spectral  functions in the  limit $p_{0}\to 0$ are
\begin{eqnarray}
&&\rho_{L}(P)\approx\frac{p_{0}m_{D}^{2}}{2p}\frac{1}{(p^{2}+m_{D}^{2})^{2}}\notag\;,\\
&&\rho_{T}(P)\approx\frac{p_{0}pm_{D}^{2}}{4}\frac{1}{p^{6}+\frac{1}{16}\pi^{2}m_{D}^{4}(p_{0})^{2}}\;,
\end{eqnarray}
where we  find  no static magnetic screening.

With above equations, we can calculate the imaginary part of $\mathrm{Tr}[\cancel{Q}\Sigma_{R}(Q)]$,
\begin{eqnarray}
&&\mathrm{Im}\Big[\mathrm{Tr}[\cancel{Q}\Sigma_{R}(Q)]\Big]\notag\\
&&=-4\pi^{2}C_{F}g^{2}\int\frac{\mathrm{d^{4}}P}{(2\pi)^{4}}\Big[4q_{\perp}^{2}\rho_{T}(P)+4\frac{(n\cdot Q)^{2}p^{2}}{(n\cdot P)^{2}}\rho_{L}(P)\Big]n_{B}(p_{0})\mathrm{sgn}(q_{0}-p_{0})\delta[(Q-P)^{2}]\notag\\
&&=-\frac{1}{2\pi}C_{F}g^{2}Tq[1+2\ln\frac{1}{g}]\;.
\end{eqnarray}

The damping rate for the hard quark is
\begin{eqnarray}
&&\Gamma_{q}(Q)=-\frac{1}{2q}\mathrm{Im}\big[\mathrm{Tr}[\cancel{Q}\Sigma_{R}(Q)]\big]\notag\\
&&=\frac{1}{4\pi}C_{F}g^{2}T[1+2\ln\frac{1}{g}]\;.
\end{eqnarray}
 The first term in the bracket comes from the longitudinal contribution of  the  HTL resumed gluon propagator,
 and the second term comes from the transverse contribution.
 In the transverse part there is an IR-cutoff,
 which results from the magnetic mass of the order $m_{magn}\sim g^{2}T$.

\begin{figure}
  \centering
  \includegraphics[width=8cm]{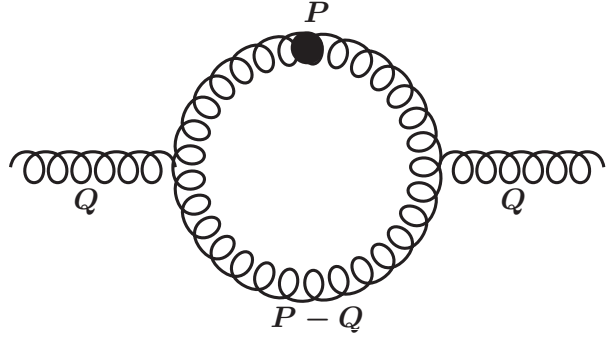}\\
  \caption{The  gluon loop diagram gives the leading order contribution to the damping rate of the hard gluon in the light cone gauge,
  and the quark loop diagram is suppressed due to the Fermi-Dirac distribution function.}\label{gluondamping}
\end{figure}

We can express the gluon loop in Fig.\ref{gluondamping} as
\begin{eqnarray}
\Pi_{\mu\alpha}^{ab}(Q)\!&=&\!\frac{i}{2}\int\frac{d^{4}P}{(2\pi)^{4}}V_{\mu\nu\rho}(Q,\!-\!P,P\!-\!Q)iG^{\nu\beta}(P)V_{\beta\gamma\alpha}(P,Q\!-\!P,\!-\!Q)id^{\rho\gamma}(P\!-\!Q)G(P\!-\!Q)\;,
\end{eqnarray}
where $G^{\nu\beta}(P)$ is the HTL resumed gluon propagator in the light cone gauge in Eq.\eqref{HTL-resummed-g-propagator},
$d^{\rho\gamma}(P\!-\!Q)G(P\!-\!Q)$ is the bare gluon propagator in the light cone gauge,
and the three gluon vertexes $V_{\mu\nu\rho}(Q,\!-\!P,P\!-\!Q)$ and $V_{\beta\gamma\alpha}(P,Q\!-\!P,\!-\!Q)$ are given in Eq.\eqref{three-g-vertexes}.

Using the relation of the Keldysh representation in \eqref{RTFGreen-functions}, we have
\begin{eqnarray}
&&\Delta_{11}^{\nu\beta}(P)\Delta_{11}(Q-P)-\Delta_{12}^{\nu\beta}(P)\Delta_{21}(Q-P)\notag\\
&&=\frac{1}{2}[\Delta_{R}^{\nu\beta}\Delta_{S}+\Delta_{S}^{\nu\beta}\Delta_{A}+\Delta_{A}^{\nu\beta}\Delta_{A}+\Delta_{R}^{\nu\beta}\Delta_{R}]\;.
\end{eqnarray}

Similarly, by power counting the second term $\Delta_{S}^{\nu\beta}\Delta_{A}$ gives the leading order contribution $O(1/g^{3})$ at finite temperature.  $\Delta_{A}^{\nu\beta}\Delta_{A}$ and $\Delta_{R}^{\nu\beta}\Delta_{R}$ are both $O(1/g^{2})$.
The internal momentum $P$ is soft, $p_{0}\sim gT$.  The Bose-Einstein distribution function $n_{B}(p_{0})$ in the term $\Delta_{S}^{\nu\beta}\Delta_{A}$ is $O(1/g)$,
\begin{equation}
\frac{1}{e^{\frac{|p_{0}|}{T}}-1}\sim\frac{T}{|p_{0}|}\propto \frac{1}{g}\;.
\end{equation}

However, the  Bose-Einstein  distribution function $n_{B}(p_{0}-q_{0})
 $ in the term $\Delta_{R}^{\nu\beta}\Delta_{S}$ is $O(1)$. The external momentum $Q$ is hard, $q_{0}\sim T$,
\begin{equation}
\frac{1}{e^{\frac{|p_{0}-q_{0}|}{T}}-1}\sim O(1)\;.
\end{equation}

The three gluon vertexes are simplified into
\begin{eqnarray}
&&V_{\mu\nu\rho}(Q,\!-\!P,P\!-\!Q)\approx gf^{acd}[Q_{\rho}g_{\mu\nu}+Q_{\mu}g_{\nu\rho}-2Q_{\nu}g_{\rho\mu}]\notag\;,\\
&&V_{\beta\gamma\alpha}(P,Q\!-\!P,\!-\!Q)\approx gf^{cdb}[-Q_{\alpha}g_{\beta\gamma}+2Q_{\beta}g_{\gamma\alpha}-Q_{\gamma}g_{\alpha\beta}]\;,
\end{eqnarray}
where the external momentum $Q$ is hard, and the internal momentum $P$ is soft, which is ignored in the above expression.

The nominator of the bare propagator in the light cone gauge $d^{\rho\gamma}(P-Q)$ is
\begin{equation}
d^{\rho\gamma}(P-Q)=-g^{\rho\gamma}+\frac{n^{\rho}(P-Q)^{\gamma}+n^{\gamma}(P-Q)^{\rho}}{n\cdot(P-Q)}\approx -g^{\rho\gamma}+\frac{n^{\rho}Q{}^{\gamma}+n^{\gamma}Q{}^{\rho}}{n\cdot Q}\;.
\end{equation}

In the calculation, we have
\begin{equation}
Q_{\rho}d^{\rho\gamma}(P-Q)=\frac{n^{\gamma}Q^{2}}{n\cdot Q}=0\;,
\end{equation}
where the external hard momentum $Q$ is on shell, $Q^{2}=0$. Due to the equation, we can simplify the calculation.

 Now we calculate the imaginary part of the transverse part $\Pi_{T}(Q)$,
\begin{eqnarray}
&&\mathrm{Im}\Pi_{T}(Q)=\frac{1}{2}(\delta^{ij}-\frac{q^{i}q^{j}}{q^{2}})\mathrm{Im}\Pi^{ij}(Q)\hspace{2cm}(i,j=1,2,3)\notag\\
&&=-C_{A}g^{2}\pi^{2}\int\frac{\mathrm{d^{4}}P}{(2\pi)^{4}}\Big[4q_{\perp}^{2}\rho_{T}(P)+4\frac{(n\cdot Q)^{2}p^{2}}{(n\cdot P)^{2}}\rho_{L}(P)\Big]n_{B}(p_{0})\mathrm{sgn}(q_{0}-p_{0})\delta[(Q-P)^{2}]\notag\\
&&=-\frac{1}{8\pi}C_{A}g^{2}Tq[1+2\ln\frac{1}{g}]\;.
\end{eqnarray}

The damping rate for the hard gluon is
\begin{equation}
\Gamma_{g}(Q)=-\frac{1}{2q}\mathrm{Im[\Pi_{T}(Q)]}=\frac{1}{16\pi}C_{A}g^{2}T
[1+2\ln\frac{1}{g}]\;.
\end{equation}
 The first term in the bracket stems from the longitudinal contribution of  the  HTL resumed gluon propagator,
 and the second term stems from the transverse contribution. For the transverse part we take an IR-cutoff too.

 We work out the result about the damping rates of the hard quark and gluon in the  limit $p_{0}\to 0$.
Below we can prove the general case that the expression about the transverse and longitudinal projection  in the above calculation of the damping rates is the same in the two gauges.

Via this function $\delta[(Q-P)^{2}]$, we have the relation about $\cos\theta$,
\begin{equation}
\cos\theta\approx\frac{p_{0}}{p}=\frac{p_{0}}{|p_{3}|}\;,
\end{equation}
where the internal soft momentum $P^{\mu}=(p_{0},0,0,p_{3})$, the angle $\theta$ is arbitrary.

When $p_{3}>0$, we have
\begin{equation}
\cos\theta=\frac{q_{3}}{q}=\frac{q_{3}}{q_{0}}\;,
\end{equation}
where $\theta$ is the angle between  $\overrightarrow{p}$ and $\overrightarrow{q}$.

The expression of the longitudinal part in the calculation of the damping rate is
\begin{equation}
\frac{(n\cdot Q)^{2}}{(n\cdot P)^{2}}P^{2}=\frac{(q_{0}+q_{3})^{2}}{(p_{0}+p_{3})^{2}}P^{2}=\frac{(q_{0})^{2}(1+\cos\theta)^{2}}{(p_{3})^{2}(1+\cos\theta)^{2}}P^{2}=\frac{(q_{0})^{2}}{(p_{3})^{2}}P^{2}=\frac{(q_{0})^{2}}{p^{2}}P^{2}\;.
\end{equation}

When $p_{3}<0$, we have
\begin{equation}
\cos\theta=-\frac{q_{3}}{q}=-\frac{q_{3}}{q_{0}}\;.
\end{equation}

The expression of the longitudinal part in the calculation is
\begin{equation}
\frac{(n\cdot Q)^{2}}{(n\cdot P)^{2}}P^{2}=\frac{(q_{0}+q_{3})^{2}}{(p_{0}+p_{3})^{2}}P^{2}=\frac{(q_{0})^{2}(1-\cos\theta)^{2}}{(p_{3})^{2}(1-\cos\theta)^{2}}P^{2}=\frac{(q_{0})^{2}}{(p_{3})^{2}}P^{2}=\frac{(q_{0})^{2}}{p^{2}}P^{2}\;.
\end{equation}

We can find, in the calculation  we have the same expression about the longitudinal projection in the two gauges.

The transverse projection tensor of the HTL resumed gluon propagator in the Coulomb gauge is
\begin{equation}
\delta^{ij}-\frac{p^{i}p^{j}}{p^{2}}\;.
 \end{equation}
By calculating, we have
\begin{equation}
\delta^{3j}-\frac{p^{3}p^{j}}{p^{2}}=0\;,
 \end{equation}
where the internal soft momentum $P^{\mu}=(p_{0},0,0,p_{3})$. So we can get the same transverse projection tensor in the two gauges.

We have the simple expression about the transverse projection tensor,
\begin{equation}
\delta^{ij},\hspace{0.5cm}(i,j=1,2).
 \end{equation}

Here we show the transverse and longitudinal projection tensors in the above calculation of the damping rates are the same in the two gauges when the transverse momentum of the internal momentum $P$ is zero.
So we can get the same result about the damping rates of the hard quark and gluon in the two gauges.

\section{Conclusion}

In this paper, we have derived the HTL  resumed gluon propagator in the light cone gauge,
and presented the results for the transverse and longitudinal  gluon HTL  self energy in the light cone gauge.
We show the quark HTL energy is independent of the light cone gauge, and get the same HTL resumed quark propagator.

Although the longitudinal and transverse expression of the gluon HTL self energy both have the light cone terms with $1/(n\cdot K)$,
there is no divergence in the transverse and longitudinal  parts, which are different  from the case at zero temperature.
By calculating,
we can find we obtain the same transverse and longitudinal gluon HTL self energies in  the light cone gauge and the Coulomb gauge,
although in the two gauges we have different transverse and longitudinal projection tensors.
We think the axial vector $n_{l}^{\mu}=(\frac{\sqrt{2}}{2},0,0,-\frac{\sqrt{2}}{2})$ have the longitudinal part and is rotated with respect to the axial vector $n_{c}^{\mu}=(1,0,0,0)$ in the Coulomb gauge,
which brings about changes.
Correspondingly, in the static limit,
we compare the component $\Pi^{00}(P)$ of the gluon HTL self energy in the light cone gauge with $\Pi^{00}(P)$ in the Coulomb gauge,
and find the result in the light cone gauge is the same as that in the Coulomb gauge.

We show    the transverse and longitudinal spectral functions of the HTL resumed gluon propagator in the light cone gauge and  the transverse and longitudinal dispersion relation.
However, in the light cone gauge,
the transverse and longitudinal projection tensors are both based on the the axial vector $n_{l}^{\mu}$,
and they have  different expression from the transverse and longitudinal projection tensors in the Coulomb gauge, so the expression of  the HTL resumed gluon propagator in the light cone gauge is different from  that  in the Coulomb gauge.

With the HTL  resumed gluon propagator in the light cone gauge, we  calculate the damping rates of the hard on shell quark and gluon in a particular limit.
We demonstrate in  general case, we can get the same result about the damping rates in the two gauges. Although  the expression of the the HTL  resumed gluon propagator in the light cone gauge is different from that in the Coulomb gauge,  we can find it is gauge independent. Using the propagator, we can further consider the correction from the soft process for some physical quantities at high temperature in the Heavy Ion Collisions.

\section{Acknowledge}

We are grateful to Xin-nian Wang for helpful discussion.
This work is supported by Ministry of Science and Technology
of China (MSTC) under "973" Project No. 2015CB856904(4)
and by NSFC under Grants No. 11375070, No. 11221504, and No. 11135011.

\section{Appendix}
At zero temperature, this kind of the  integral in Eq.\eqref{light-cone-termT} has  divergence,
which has been renormalized successfully.
However, we find this integral does not have  divergence in the HTL approximation.

\begin{figure}
  \centering
  \includegraphics[width=5cm]{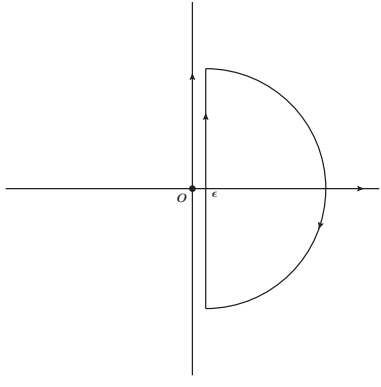}\\
  \caption{The contour integral for boson.}\label{contour-integral}
\end{figure}

We use the contour integral to substitute  the frequency sum of boson at finite temperature \cite{K1989},
\begin{eqnarray}\label{integral contour}
T\stackrel[n=-\infty]{\infty}{\sum}f(p_{0}=2n\pi Ti)=\frac{1}{2\pi i}\int_{-i\infty}^{i\infty}\mathrm{d}p_{0}\frac{1}{2}[f(p_{0})+f(-p_{0})]+\frac{1}{2\pi i}\int_{-i\infty+\epsilon}^{i\infty+\epsilon}\mathrm{d}p_{0}[f(p_{0})+f(-p_{0})]\frac{1}{e^{\beta p_{0}}-1}\;.
\end{eqnarray}
This equation has the zero temperature part and  the finite temperature part, and we only consider the the finite temperature part here.

 Below we prove this integral at finite temperature does not have the divergence in Eq.\eqref{light-cone-termT},
\begin{equation}
\frac{1}{2\pi i}\int_{-i\infty+\epsilon}^{i\infty+\epsilon}\mathrm{d}k_{0}\int\frac{\mathrm{d^{3}\overrightarrow{k}}}{(2\pi)^{3}}
 \frac{K^{2}-2K\cdot P}{n\cdot KK^{2}(K-P)^{2}}\;.
\end{equation}
where $n\cdot K=n_{0}k_{0}-\overrightarrow{n}\overrightarrow{k}=n(k_{0}+k\cos\theta)$,
the angle $\theta$ is the angle between $\overrightarrow{k}$ and $\overrightarrow{p}$,
and the axial vector $n_{l}^{\mu}=(\frac{\sqrt{2}}{2},0,0,-\frac{\sqrt{2}}{2})$.

When $\cos\theta>0$,
use the equation of the the integral contour in Eq.\eqref{integral contour}, we have
\begin{small}
\begin{eqnarray}
\frac{1}{2\pi i}\int_{-i\infty\!+\!\epsilon}^{i\infty\!+\!\epsilon}\mathrm{d}k_{0}\int\frac{\mathrm{d^{3}\overrightarrow{k}}}{(2\pi)^{3}}
 \Big[\frac{K^{2}\!-\!2K\cdot P}{n(k_{0}\!+\!k\cos\theta)K^{2}(K\!-\!P)^{2}}\!+\!\frac{K^{2}\!-\!2(-k_{0}p_{0}\!-\!kp\cos\theta)}{n(-k_{0}\!+\!k\cos\theta)K^{2}[K^{2}\!-\!2(-k_{0}p_{0}\!-\!2kp\cos\theta)\!+\!P^{2}]}\Big]\frac{1}{e^{\beta k_{0}}\!-\!1}\;.
\end{eqnarray}\end{small}

The first term in the bracket has poles at $k_{0}=k$ and $k_{0}=p_{0}+|\overrightarrow{k}-\overrightarrow{p}|$,
and the residues of the two poles do not have the term of $\frac{1}{\cos\theta-1}$,
so there is no divergence at $\cos\theta=1$.
The second term has poles at $k_{0}=k$,
$k_{0}=k\cos\theta$ and $k_{0}=-p_{0}+|\overrightarrow{k}-\overrightarrow{p}|$.
The residues with the two poles of $k_{0}=k$ and $k_{0}=k\cos\theta$ contain the terms of $\frac{1}{\cos\theta-1}$,
but the sum of the two residues is finite when $\cos\theta\to 1$,
\begin{equation}
\underset{\cos\theta\to1}{\lim}\frac{1}{(-k+k\cos\theta)2k}-\frac{1}{k^{2}(\cos^{2}\theta-1)}
 =\frac{1}{2k^{2}(\cos\theta+1)}\;.
\end{equation}

When $\cos\theta<0$, we have

\begin{small}
\begin{eqnarray}
\frac{1}{2\pi i}\int_{-i\infty\!+\!\epsilon}^{i\infty\!+\!\epsilon}\mathrm{d}k_{0}\int\frac{\mathrm{d^{3}\overrightarrow{k}}}{(2\pi)^{3}}
 \Big[\frac{K^{2}\!-\!2K\cdot P}{n(k_{0}\!+\!k\cos\theta)K^{2}(K\!-\!P)^{2}}\!+\!\frac{K^{2}\!-\!2(-k_{0}p_{0}\!-\!kp\cos\theta)}{n(-k_{0}\!+\!k\cos\theta)K^{2}[K^{2}\!-\!2(-k_{0}p_{0}\!-\!2kp\cos\theta)\!+\!P^{2}]}\Big]\frac{1}{e^{\beta k_{0}}\!-\!1}\;.
\end{eqnarray}\end{small}

The second term in the bracket has poles at $k_{0}=k$ and $k_{0}=-p_{0}+|\overrightarrow{k}-\overrightarrow{p}|$,  and the residues of the two poles do not have the term of $\frac{1}{\cos\theta+1}$, so there is no divergence at $\cos\theta=-1$. The first term has poles at $k_{0}=k$, $k_{0}=-k\cos\theta$ and $k_{0}=p_{0}+|\overrightarrow{k}-\overrightarrow{p}|$.
The residues with the two poles of $k_{0}=k$ and $k_{0}=-k\cos\theta$ contain the terms of $\frac{1}{\cos\theta+1}$, but the sum of the two residues is finite when $\cos\theta\to -1$,
\begin{equation}
\underset{\cos\theta\to-1}{\lim}\frac{1}{k(1+\cos\theta)2k}+\frac{1}{k^{2}(\cos^{2}\theta-1)}=\underset{\cos\theta\to-1}{\lim}\frac{1}{2k^{2}(\cos\theta-1)}\;.
\end{equation}

Combining the situation $\cos\theta>0$ and $\cos\theta<0$, the integral at finite temperature in Eq.\eqref{light-cone-termT} does not have divergence.
The integral is $O(g^{3}T^{2})$, which can be ignored.


\end{document}